\documentclass[a4paper]{JACoW-GSI}
\usepackage{graphicx,multicol}
\usepackage{url}
\usepackage{hyperref}

\begin{document}
\title{Bayesian Analysis of Hybrid EoS based on Astrophysical Observational Data}

\author{D.E.~Alvarez-Castillo$\sp{1,2}$}
\author{A.~Ayriyan$\sp{3,4}$}
\author{D.~Blaschke$\sp{1,5}$}
\author{H.~Grigorian$\sp{3,6}$}
\affil{$\sp{1}$Bogoliubov Laboratory of Theoretical Physics, JINR, Russia}
\affil{$\sp{2}$Instituto de Fisica, Universidad Autonoma de San Luis Potosi, Mexico}
\affil{$\sp{3}$Laboratory of Information Technologies, JINR, Russia}
\affil{$\sp{4}$Tver State University, Russia}
\affil{$\sp{5}$Institute of Theoretical Physics, University of Wroclaw, Poland}
\affil{$\sp{6}$Yerevan State University, Armenia}
\maketitle

\begin{abstract}
We perform a Bayesian analysis of probability measures for compact star
equations of state using new, disjunct constraints for mass and radius. 
The analysis uses a simple parametrization for hybrid equations of state 
to investigate the possibility of a first order deconfinement transition 
in compact stars. 
The latter question is relevant for the possible existence of a critical 
endpoint in the QCD phase diagram under scrutiny in heavy-ion collisions. 
\end{abstract}

\section{Introduction}
The most basic features of neutron stars (NS) are their radii and masses which 
so far have not been well determined simultaneously for a single object. 
In some cases masses are precisely measured like in the case of binary systems 
but radii are quite uncertain. 
In the other hand, for isolated NS some radius and mass measurements
exist but lack the necessary precision to inquire into their interiors. 
In fact, it has been conjectured that there exists a unique relationship 
for all NS between the sequence of mass and radius relations on the 
one hand and their equation of state (EoS) on the other that determines their 
internal composition \cite{Lindblom1984}. 
For this reason, accurate observations of masses and radii are crucial to 
study cold dense nuclear matter as it exists in NS.

However, the present observable data allow to make only probabilistic 
estimations of the internal structure of the star which can be performed 
using Bayesian Analysis (BA) and modeling of relativistic configurations of 
NS. 
In comparison to previous work in this direction \cite{Steiner:2010fz}
we choose disjunct mass-radius constraints in this work which reveals that
probabilistic estimations of the superdense stellar matter EoS have a much more
preliminary character at present than obtained previously. 
Moreover, we focus in this analysis on investigating the possibility of 
a quark matter inner core in massive NS with $M\sim 2~M_\odot$ 
(see \cite{Demorest:2010bx,Antoniadis2013}),
separated from the outer core of hadronic matter by a strong first order phase 
transition.

\section{NS structure}
The microscopical properties of compact stars are modeled in the framework of 
general relativity, where the Einstein equations are solved for
a static (non-rotating), spherical star resulting in the 
Tolman--Oppenheimer--Volkoff (TOV) equations 
\cite{Tolman:1939jz,Oppenheimer:1939ne}, see also \cite{Glendenning2000}

\begin{eqnarray}
\frac{d\ m(r)}{dr} &=& C_1\varepsilon(r)\ r^{2}~,
\\
\frac{d\ p(r)}{dr} &=& -C_2\frac{(\varepsilon(r)+p(r0)
	(m(r)+C_1p(r)\ r^{3})}{r(r-2C_2m(r))}~,
\label{TOV}
\end{eqnarray}
as well as the equation for the baryon mass profile $m_B(r)=m_Nn(r)$ of the 
star
\begin{equation}
\frac{dn(r)}{dr} = C_1 n(r)\frac{r^{2}}{\sqrt{1-2C_2m(r)/r}}
\end{equation}
with $m_{N}$ being the atomic mass unit and the constants are defined as
\begin{eqnarray}
C_1&=&\frac{4\pi}{c^2}=1.11269 \cdot 10^{-5}\frac{\textrm{M}_{\odot}}
{\textrm{km}^{3}}\frac{\textrm{fm}^{3}}{\textrm{MeV}}~,
\\
C_2&=&\frac{G}{c^2}=1.4766 \frac{\textrm{km}}{\textrm{M}_\odot}
~.
\label{consts}
\end{eqnarray}

These equations are integrated from the center of the star towards its surface,
with the radius of the star $R$ defined by the condition $p(R)=0$ while the 
gravitational mass is $M=m(R)$.
In a similar manner, the baryon mass is given by $M_{B}=m_{B}(R)$.

In order to solve the TOV equations the EoS is required. 
It is given by the relation $p=p(\varepsilon)$ which carries information about 
the microscopic state of dense nuclear matter. 
Thus, the above equations have to be solved simultaneously using as a boundary 
condition the energy density at the star centre $\varepsilon(r=0)$.

In this way, for a given value of $\varepsilon(0)$ the solution of the 
TOV equations are the $p(r)$ and $m(r)$ profiles and with them the  
relation $M(R)$ in the parametric form $M(\varepsilon(0))$ and 
$R(\varepsilon(0))$.

\section{The hybrid EoS}
For this study we follow the AHP scheme~\cite{Alford2013} for defining the 
hybrid EoS in the form
\begin{eqnarray}
p(\varepsilon)&=& p_h(\varepsilon)\Theta(\varepsilon_H-\varepsilon)
+ p_q(\varepsilon)\Theta(\varepsilon-\varepsilon_Q),
\nonumber\\ 
&& +p_h(\varepsilon_H)\Theta(\varepsilon-\varepsilon_H)
\Theta(\varepsilon_Q -\varepsilon)
\end{eqnarray}
where $p_h$ is the pressure of a pure hadronic EoS and $p_q$ represents the 
high density EoS assumed here as quark matter with $c^{2}_q$, its 
squared speed of sound, as parametrized by Haensel et al.~\cite{Zdunik2013}  
which describes pretty well the superconducting NJL model derived 
in~\cite{Blaschke:2005uj,Klahn:2006iw,Klahn:2013kga}. 
For the hadronic EoS we take the well known model of APR~\cite{Akmal1998} that 
is in agreement with experimental data of densities about nuclear saturation. 
For this hadronic branch all the relevant thermodynamical variables: 
energy density $\varepsilon$, pressure $p$, baryon density $n$ and chemical 
potential $\mu$ are well defined and taken as input for determination of 
the hybrid (hadronic + quark matter) EoS. 
As a starting point in the derivation of the high density EoS we introduce the 
pressure as function of energy density in the quark matter phase
\begin{equation}
p_q(\varepsilon)=c^{2}_q\varepsilon-B,
\end{equation}
with $B$ playing the role of a bag constant. 
To determine the remaining thermodynamical quantities $n_q$ and $\mu_q$ of the 
quark matter phase we use the following relations
\begin{eqnarray}
 n_q(\varepsilon)&=&n_Q\exp\left(\int_{\varepsilon_Q}^{\varepsilon}
     \frac{d\varepsilon'}{\varepsilon'+p(\varepsilon')}\right)\\
 \mu&=&\frac{\epsilon+p}{n}
\end{eqnarray}
where $\varepsilon_Q$ is the quark matter energy density right after the phase 
transition following the jump $\Delta \epsilon$ as density increases. 
Therefore one arrives at the following formula
\begin{equation}
n_q(\varepsilon)=n_Q~\frac{p_c+\varepsilon_Q}{p_c+\varepsilon_H}~
\left(\frac{(1+c^{2}_q)\varepsilon-B}{(1+c^{2}_q)\varepsilon_Q-B}\right)^{\frac{1}{1+c^{2}_q}},
\end{equation}
obtained by enforcing conditions of equal pressure and chemical potential at 
the transition (Gibbs conditions):
\begin{equation}
 \mu_c=\frac{p_c+\varepsilon_Q}{n_Q}=\frac{p_c+\varepsilon_H}{n_H},
\end{equation}
where upper case subscripts stand for the values of the thermodynamic 
functions at the phase transition:
$\varepsilon_Q=\varepsilon_q(\mu_c)$, $\varepsilon_H=\varepsilon_h(\mu_c)$,
$n_Q=n_q(\mu_c)$, $n_H=n_h(\mu_c)$, etc.

\begin{figure}[ht!]
\begin{center}$
\begin{array}{c}
\includegraphics[width=7cm]{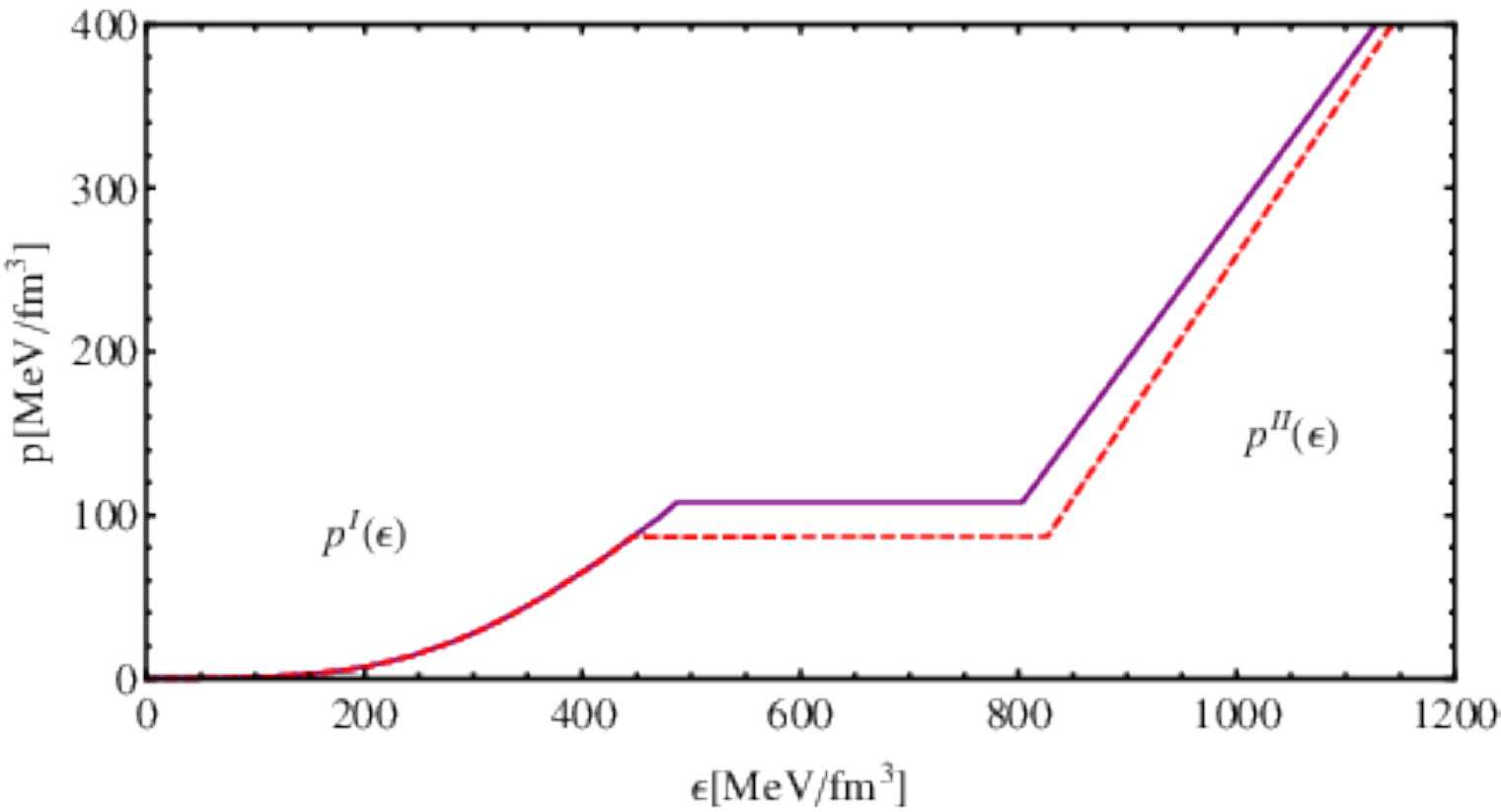}\\
\includegraphics[width=7cm]{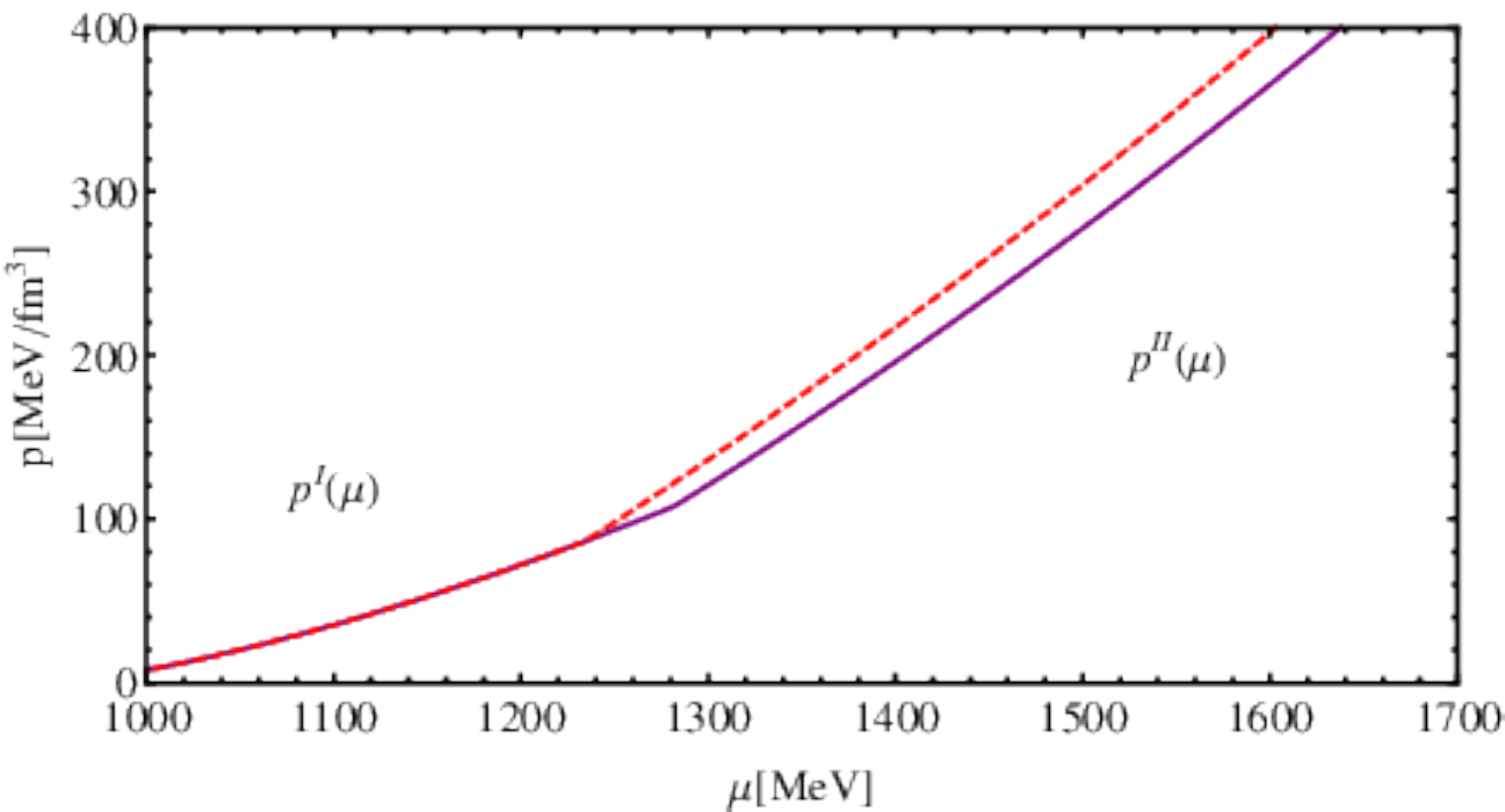}
\end{array}$
\end{center}
\caption{Hybrid EoS scheme for two different sets of the three parameters 
$\left(\epsilon_H,\gamma,{c}_{q}^{2}\right)$.}
\label{APH_Scheme}
\end{figure}

The free parametes of the model are the transition density $\varepsilon_H$, 
the energy density jump $\Delta \varepsilon \equiv \gamma \varepsilon_H$ and 
$c^{2}_q$, the quark matter speed of sound squared. 
The resulting EoS in the plane pressure versus density is depicted in 
Fig.~\ref{APH_Scheme} for a given set of input parameters.

\section{Bayesian Analysis Formulation}
We define the vector of free parameters 
$\overrightarrow{\pi}(\varepsilon_H,\gamma,c^{2}_q)$, 
which define the EoS with phase transition from nuclear to quark matter. 
The nuclear equation of state can be taken to be~APR~\cite{Akmal1998}.

These parameters were sampled as
\begin{equation}
\label{pi_vec}
\pi_i = \overrightarrow{\pi}
\left(\varepsilon_H(k),\gamma(l),{{c}_{q}^{2}}(m)\right),
\end{equation}
where $i = 0\dots N-1$ with $N = N_1\times N_2\times N_3$ 
such that $i = N_1\times N_2\times k + N_2\times l + m$ and 
$k = 0\dots N_1-1$, $l = 0\dots N_2-1$, $m = 0\dots N_3-1$, here $N_1$, $N_2$ and $N_3$ number of parameters $\epsilon_k$, $\gamma_l$ and ${{c}_{s}^{2}}_{m}$ respectively.

Using the EoS one can calculate the neutron star structure by solving the TOV 
equations. Then it is possible to use different neutron star observations to 
check the probability for this EoS. 
We use here three observations as constraints: 
a mass constraint~\cite{Antoniadis2013}, 
a radius constraint~\cite{Bogdanov2013} and 
a constraint for the gravitational binding energy (relation between 
gravtitational mass and baryon mass)~\cite{Podsiadlowski:2005ig,Kitaura2006}.

The goal is to find the set of the most probable $\pi_i$ based on the given 
constraints using Bayesian Analysis (BA).
For initializing the BA we propose that {\it a priori} each vector of 
parameters $\pi_i$ has a probability equal to one, $P\left(\pi_i\right) = 1$, 
for all $i$. 

\subsection{Mass Constraint} 
We assume that error of the mass measurement for the high-mass pulsar 
PSR~J0348+0432~\cite{Antoniadis2013} is normal distributed with
$\mathcal{N}(\mu_A,\sigma_A^2)$, where the mean value is 
$\mu_A = 2.01~\mathrm{M_{\odot}}$ and the variance is
$\sigma_A = 0.04~\mathrm{M_{\odot}}$. 
Using this assumption we can calculate the conditional probability of the 
event $E_{A}$ that the mass of a neutron star corresponds to measurement:
\begin{equation}
\label{p_anton}
P\left(E_{A}\left|\pi_i\right.\right) = \Phi(M_i, \mu_A, \sigma_A),
\end{equation}
where $M_i$ - maximal mass constructed by $\pi_i$ and $\Phi(x, \mu, \sigma)$ 
is the cumulative distribution function for the normal distribution
\begin{equation}
\label{Laplas}
\Phi(x, \mu, \sigma) = 
\frac{1}{2}
\left[1+{\rm erf}\left(\frac{x-\mu}{\sqrt{2\sigma^2}}\right)\right].
\end{equation}

\subsection{Radius Constraint}
In the BA by Steiner et al. \cite{Steiner:2010fz} the luminosity radius 
extracted for burst sources has been used to  constrain a combined mass-radius 
relationship. 
This method is problematic, in particular because of the unknown stellar 
atmosphere composition, uncertainties in the distance to the 
source, the bias of the parabolic M-R constraint with the shape of stellar
sequences in the M-R diagram for typical EoS and last not least due to unknown 
details of the burst mechanism.     
A very promising technique to measure radii of neutron stars  is based on the 
analysis of pulsar timing residuals since it does not rely on knowledge of the
luminosity of thermal radiation. 
Such a radius measurement gives $\mu_B = 15.5~\mathrm{km}$ and 
$\sigma_B = 1.5~\mathrm{km}$ for PSR~J0437-4715~\cite{Bogdanov2013}. 
A similar range of radius values has recently been obtained for RXJ 1856 by
Hambaryan et al. 
Now it is possible to calculate the conditional probability of the event 
$E_{B}$ that the radius of neutron star corresponds to the given measurement
\begin{equation}
\label{p_bogdan}
P\left(E_{B}\left|\pi_i\right.\right) = \Phi(R_i, \mu_B, \sigma_B)~.
\end{equation}

\subsection{$M_G$--$M_B$ Relation Constraint}
This constraint corresponds to a region in the $M_G$--$M_B$ plane. 
We need to estimate the probability that a  point 
$M_i = \left({M_G}_i, {M_B}_i\right)$ is close to the point 
$\mu = \left(\mu_G, \mu_B\right)$. 
The mean values $\mu_G = 1.249$, $\mu_B = 1.36$ and standard deviations 
$\sigma_{M_G} = 0.001$, $\sigma_{M_B} = 0.002$ are given in 
\cite{Kitaura2006}. 
The needed probability can be calculated by the formula
\begin{equation}
\label{p_kitaura}
\small
P\left(E_{K}\left|\pi_i\right.\right) = 
\left[ \Phi\left(\xi_G\right) - \Phi\left(-\xi_G\right) \right]
\left[ \Phi\left(\xi_B\right) - \Phi\left(-\xi_B\right) \right],
\end{equation}
where $\Phi\left(x\right) = \Phi\left(x, 0, 1\right)$, 
$\xi_G = {\sigma_{M_G}}/{d_{M_G}}$ and 
$\xi_B = {\sigma_{M_B}}/{d_{M_B}}$, 
$d_{M_G}$ and $d_{M_B}$ are the absolute values of the components of the vector
$\mathrm{\textbf{d}} = \mathrm{\bf\mu} - \mathrm{\textbf{M}}_i$. 
Here $\mathrm{\bf\mu} = \left(\mu_G, \mu_B\right)^T$ is given in 
\cite{Kitaura2006} and 
$\mathrm{\textbf{M}}_i = \left({M_G}_i, {M_B}_i\right)^T$ is the solution of 
the TOV equations using the $i^{\mathrm{th}}$ vector of EoS parameters $\pi_i$.
Note that formula (\ref{p_kitaura}) does not correspond to a multivariate 
normal distribution.

\subsection{Calculation of {\it a posteriori} Probabilities}
Note, that these measurements are independent of each other. 
This means that we can calculate the complete conditional probability of an 
event $E$ given $\pi_i$ corresponds to the product of the conditional 
probabilities of all measurements, in our case resulting from the three 
constraints $E_A$, $E_B$, $E_K$, 
\begin{equation}
\label{p_event}
P\left(E\left|\pi_i\right.\right) = 
P\left(E_{A}\left|\pi_i\right.\right) 
\cdot P\left(E_{B}\left|\pi_i\right.\right) 
\cdot P\left(E_{K}\left|\pi_i\right.\right).
\end{equation}
Now, we can calculate the probability of $\pi_i$ using Bayes' theorem:
\begin{equation}
\label{pi_apost}
P\left(\pi_i\left|E\right.\right) = 
\frac{P\left(E\left|\pi_i\right.\right)
P\left(\pi_i\right)}{\sum\limits_{j=0}^{N-1}P\left(E\left|\pi_j\right.\right)P\left(\pi_j\right)}.
\end{equation}
\begin{figure}[ht!]
\begin{center}$
\begin{array}{c}
\includegraphics[width=0.5\textwidth]{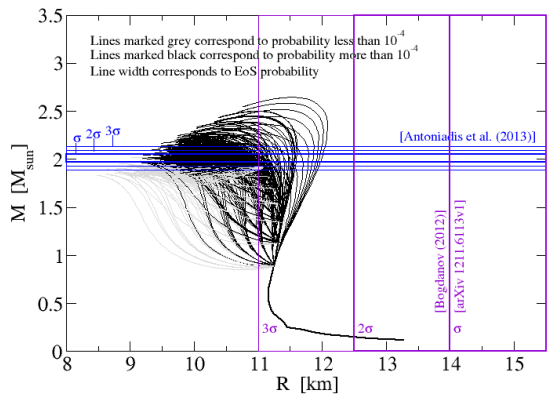}
\\
\includegraphics[width=0.5\textwidth]{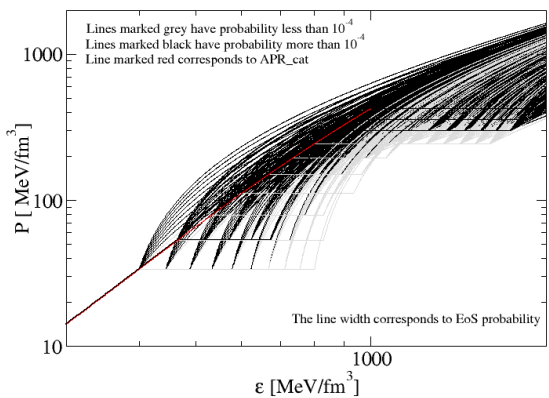}
\end{array}$
\end{center}
\caption{Mass-radius relations (upper panel) for different sets of NS 
configurations corresponding to pressure vs. energy density relations
(lower panel) which are obtained by varying three EoS parameters, see text.
The thickness of the lines is proportional to the probability value for the 
parameter vector $\pi$.}
\label{results}
\end{figure}

\section{Results and Discussion}

We apply the scheme of BA for the probabilistic estimation of the EoS given by 
the vector parameter $\pi$. 
Varying the parameters 
in the intervals $400<\varepsilon_H[\mathrm{MeV/fm^{3}}] <1000$, 
$0<\gamma<1$ and $0.3<{c}_{q}^{2}<1$ we explore the calculations for
$N=10^3$ and the results are presented in Fig.~\ref{results}. 
The mass-radius relation is shown in the upper panel and the pressure as a 
function of the energy density is in the lower one. 
The results are shown for different sets of neutron star configurations 
corresponding to different sets of EoS parameters. 
The thickness of the lines is chosen to be proportional to the probability 
value for the parameter vector $\pi$.
We~show that the chosen constraints are not sufficient to distinguish two 
cases: the one with the existence of a third family of twin stars 
(two stars with same masses but different radii due to different internal 
composition) from the case where only neutron star family is possible.
This result can also be obtained from the lower panel of Fig.~\ref{results} 
where the EoS are plotted with different line thickness corresponding to the 
probability value for the parameter vector $\pi$.
It is apparent that the two types of EoS, with and without the phase 
transition to quark matter, have approximately the same probability.
Nevertheless, when the phase transition to quark matter is possible then our 
constraint requires that the phase transition should occur for energy 
densities exceeding $900~\mathrm{MeV/fm^{3}}$ with a jump in energy density
up to $10^3~\mathrm{MeV/fm^{3}}$.

We conclude that the current state of knowledge of observables for masses and 
radii of compact stars does not yet allow to extract with certainty a statement
about the possible existence of a quark matter inner core with a strong first 
order phase transition to the outer core.

\section{Acknowledgments}
The research was supported in part by the Bogoliubov-Infeld Program for 
collaboration between JINR Dubna and Polish Universities and Institutes, the
Ter-Antonian-Smorodinsky Program for collaboration between JINR Dubna and 
Armenian Universities and Institutes. 
A.~Ayriyan acknowledges support by JINR under grant No. 14-602-01.


\end{document}